\documentclass[twocolumn,prb,showpacs,preprintnumbers,amsmath,amssymb]{revtex4}
\usepackage{graphicx}
\usepackage{dcolumn}
\input epsf

\begin{document}

\title{Spin susceptibility of underdoped cuprates: the case of Ortho-II YBa$_2$Cu$_3$O$_{6.5}$}
\author{E. Bascones$^{1,2}$, T. M. Rice$^2$}
\affiliation{$^1$Instituto de Ciencia de Materiales de Madrid, CSIC,Cantoblanco, E-28049 Madrid, Spain \\ $^2$Theoretische Physik, ETH-H\"onggerberg, CH-8050 Zurich, Switzerland }
\date{\today}
\begin{abstract}
Recent inelastic neutron scattering measurements found that the spin 
susceptibility of detwinned and highly ordered  ortho-II 
YBa$_2$Cu$_3$0$_{6.5}$ exhibits, in both the normal and superconducting states, one-dimensional incommensurate modulations at 
low energies  which were interpreted as 
a signature of dynamic stripes. We propose an alternative model based on 
quasiparticle transitions between the arcs of a truncated Fermi surface. 
Such transitions are 
resonantly enhanced by scattering to the triplet spin resonance. We show that 
the anisotropy 
in the experimental spin response  is consistent with this model if the 
gap at the saddle points is anisotropic.
\end{abstract}
\email{leni@icmm.csic.es}
\pacs{74.72-h,75.25+z,76.50+g}
\maketitle

Recent experiments\cite{Stock02,Stock04,Stock05} have addressed the 
magnetic spectrum in detwinned and 
highly ordered ortho-II $YBa_2Cu_3O_{6.5}$\cite{Liang00}
which has an average doping per planar 
Cu $x \sim 0.09$ and $T_c \sim 60$ K. The ortho-II phase 
is characterized by a periodic alternation of filled and 
empty Cu-O $b$-axis chains, doubling the size of the unit cell in the $a$ 
direction\cite{Andersen99}. The oxygen ordering reduces the 
disorder making it an ideal candidate for the study of the magnetic properties at underdoping. The magnetic response shows a resonance\cite{Stock04} at 33 meV at 
 $Q=(\pi,\pi)$ and a {\it ring}-like high energy branch 
dispersing upwards\cite{Stock05}.
At lower energies, there is inelastic scattering from one-dimensional 
modulations at incommensurate wavevectors stretching down to zero 
energy
but without elastic or quasielastic peaks.
On entering the 
superconducting state, the scattering at energies less than 16 meV is 
{\it suppressed}, but not eliminated. 
In this strongly underdoped regime 
the incommensurability $\delta$ at low energies is very small.
Contrary to what has been observed at optimal doping the incommensurability
below the resonance persists in the normal pseudogap phase\cite{Stock04}.
The feature associated with the one-dimensional modulations, which were interpreted as a signature of dynamic stripes\cite{Stock04}, is a 
flat top lineshape  centered at $(\pi,\pi)$ along the direction
perpendicular to the chains.
An anisotropic response, but no sign of one-dimensionality has
been reported by Hinkov {\it et al}\cite{Hinkov04}  on 
untwinned 
YBa$_2$Cu$_3$O$_{6+x}$.

Static stripes should show up as elastic magnetic peaks 
and
quasi-elastic peaks would be expected\cite{Kivelsonreview}, from slow fluctuating stripes, 
but these are not observed. 
Also, the dramatic drop of the 
magnetic scattering
at low energies, at the onset of superconductivity,
is not expected if ordered antiferromagnetic regions between hole stripes
are involved. The stripe-based model also has difficulties to explain the 
isotropy found above the resonance\cite{Stock05}.

Many authors have used  RPA\cite{RPA} (random phase 
approximation) schemes to describe the magnetic 
response near optimal doping. 
This 
model relates the resonance peak and 
inconmensurate branches with
a particle-hole collective mode bounded below the superconducting gap.
An extension to the normal pseudogap phase at underdoping can be made by 
assuming an energy gap from another origin (e.g. d-density wave order\cite{Tewari}) but even then careful tuning 
of the parameters is required since the resonance decreases while the gap 
increases in energy with decreasing doping. 
Alternatively one can interpret the resonance 
as a collective mode of a spin liquid which goes soft as the antiferromagnetic 
instability at strong underdoping is approached\cite{Lee} or as an undamped spin-wave excitation brought about by strong antiferromagnetic correlations\cite{Pines}.

In this letter we introduce a simple phenomenological 
model for the  pseudogap  phase and 
focus on the anisotropic incommensurate low energy response. 
We start from a Fermi surface of quasiparticles which is truncated near the 
saddle points leaving only four Fermi arcs centered on the diagonals, similar 
to the form observed in ARPES\cite{Shenreview} experiments. Although the dispersion we use is 
derived from a staggered flux phase\cite{flux,ddw}, we propose it as a phenomenological form 
for the normal state of a doped  spin liquid (SL). Arguments that a Fermi 
surface truncation can occur without symmetry breaking have been given by 
Ledermann {\it et al}\cite{Ledermann00}, based on an analysis of multi-leg 
Hubbard ladders and by Honerkamp {\it et al}\cite{Honerkamp01} and also 
L\"auchli {\it et al}\cite{Lauchli04} from a functional renormalization group 
analysis of a 2-dimensional Hubbard model. In addition to the quasiparticle 
scattering we introduce a resonant spin triplet mode at $(\pi,\pi)$ dispersing upwards, in both normal and superconducting phases. We view the resonance
as the collective mode of the spin liquid analogous to that found, for example,
in coupled ladder systems\cite{Normand96,Tsunetsugu94,ladders} and in the slave boson approach\cite{Lee}.

We propose that the low-energy spin response is due to 
particle-hole excitations across 
the Fermi arcs. In the absence of other effects the bare spin susceptibility, due to particle-hole excitations,
is too small to explain the experimental data and has a much 
richer structure than the one observed.
In our model  such 
transitions are resonantly enhanced by scattering into the triplet spin 
resonance. 
The enhancement is strongest close to $(\pi,\pi)$.
We analyze the possible sources of anisotropy between directions parallel and
perpendicular to the chains and find that results similar to those found 
experimentally can be obtained if the gap at the saddle points is sufficiently 
anisotropic.

We start from a two-dimensional dispersion with hopping to first and second 
nearest neighbors. Truncation is introduced through 
the opening of a gap in the particle-hole channel 
at the boundary of the Antiferromagnetic Brillouin zone (AFBZ).  
Note, as argued above, we assume that this gap can occur in a  spin liquid 
without symmetry breaking. 
We neglect bilayer splitting of the bands and work with a single-planar band with dispersion:
\begin{eqnarray}
\nonumber
\epsilon_{\bf k} & = & 4 t' \cos kx \cos ky - \mu \\
& \pm & \sqrt{ 4 t^2 \left (\cos kx + \cos ky \right )^2 + \Delta^2_{SL}({\bf k})}
\end{eqnarray}
The sign follows the opposite of $(\cos kx + \cos ky)$.
The gap is assumed to have the form
$\Delta_{SL}({\bf k})=\Delta_{SP}\left ( \cos kx - \cos ky + \alpha \left [ \cos^2 kx + \cos^2 ky \right ] \right )$. If $\alpha$ is zero the gap has d-wave symmetry 
and the same magnitude at $(0,\pi)$ and $(\pi,0)$.When $\alpha$ is negative (positive) the magnitude of the gap is anisotropic and larger at $(\pi,0)$, (respectively $(0,\pi)$).
The $2a$ periodicity due to the oxygen ordering breaks the tetragonal symmetry 
and splits this band into two: $\alpha$ and $\beta$, folding the Brillouin 
zone in the $(0,0)-(\pi,0)$ direction, 
and opening a gap $V$ at $k_x=\pi/2$, see 
Bascones {\it et al}\cite{Bascones05}.In the reduced Brillouin zone, $k_x \in (-\pi/2,\pi/2)$ and $k_y \in (-\pi,\pi)$,
\begin{equation}
\epsilon^{\alpha,\beta}_{\bf k}=\frac{\epsilon_{\bf k}+\epsilon_{{\bf k}+\pi_x}}{2}\mp\frac{1}{2} {\sqrt{\left (\epsilon_{{\bf k}+\pi_x}-
\epsilon_{\bf k}\right )^2+V^2}}
\end{equation}
with  $\pi_x=(\pi,0)$.
In the superconducting state, the energy of the 
quasiparticles is $E^{\alpha,\beta}_{\bf k}=\left [ \left ( \epsilon^{\alpha,\beta}_{\bf k} \right )^2 + \left ( \Delta^{\alpha,\beta}_{\bf k} \right )^2 \right ]^{1/2}$. 
Due to the ortho-II ordering and the breaking of the tetragonal symmetry the
superconducting gap is modified from a pure d-wave form to
$\Delta^{\alpha,\beta}_{\bf k}=\frac{\Delta_0}{2}\left [g({\bf k})coskx\mp cosky\right ]$
with
\begin{equation}
g({\bf k})=\left (\epsilon_{{\bf k}+\pi_x}-\epsilon_{\bf k}\right )/\left [\left (\epsilon_{{\bf k}+\pi_x}-
\epsilon_{\bf k}\right )^2+V^2\right ]^{1/2}
\end{equation}
The resonantly enhanced\cite{resnote} spin susceptibility at low energies
is given by:
\begin{equation}
Im \chi ({\bf q},\omega)=J Re \chi^{res}({\bf q},\omega)Im\chi_0({\bf q},\omega)
\end{equation}
$J$ is a constant of order the exchange energy, 
$Im\chi_0({\bf q},\omega)$ is the imaginary part of the
bare spin susceptibility of the truncated
Fermi surface and $Re \chi^{res}({\bf q},\omega)$ is the real part arising from the resonant mode. For simplicity, we assume that the high energy part of the spin fluctuation spectrum is a  
single mode with  dispersion
taken from experiment\cite{Stock05}
\begin{equation}
\omega_{res}({\bf q})=\left (\Delta^2_{spin}+(A{\bf q})^2 \right )^{1/2}
\end{equation}
Here $\Delta^{spin}$=33 meV and $A$=365 meV $\AA$. From Kramers-Kr\"onig the real part 
\begin{equation}
Re \chi^{res}({\bf q},\omega)=1/\left[\omega -\omega_{res}({\bf q})\right]
\end{equation}
At zero temperature, applying a Kubo formula, 
\begin{eqnarray}
\nonumber
Im\chi_0({\bf q},\omega) = 
\frac{1}{2}  \sum_{{\bf k}}\left \{ 
F^\pm_{{\bf k},{\bf q}} \right.
\sum_{m,m'=\alpha,\beta}
\Lambda^{mm'}_{{\bf k},{\bf q}}  \\ 
\left.
\left [\delta
    \left ( \omega-
         \left [E^{m'}_{{\bf k}+{\bf q}}+E^m_{\bf k}
         \right ]
    \right )+\delta
    \left ( \omega+
        \left [ E^{m'}_{{\bf k}+{\bf q}}+E^m_{\bf k}
        \right ]
    \right )
  \right ] 
\right \}  
\end{eqnarray}

\begin{figure}
 \begin{center}
\leavevmode
\epsfxsize=85mm 
\epsfbox{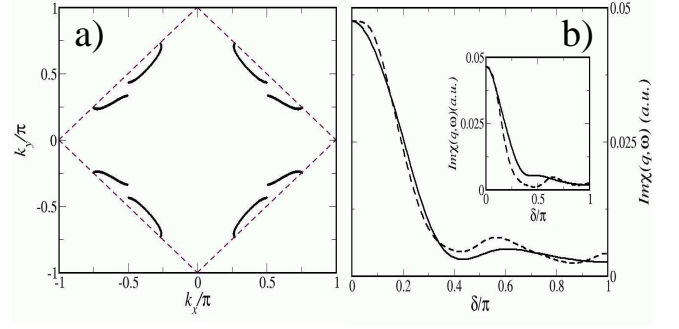}
 \end{center} 
\vspace{0mm}
\caption{a) Truncated Fermi surface, plotted in an extended Brillouin zone, 
for the band parameters discussed in the text with the tetragonal symmetry  
only broken by the inclusion of the ortho-II ordering. b) Spin susceptibility -convoluted with the resolution factor -versus momentum, measured from 
$(\pi,\pi)$ along $(\delta,0)$ and $(0,\delta)$ in solid and dashed lines, 
respectively, for the same parameters and $\omega=0.05$. The inset shows the 
same plot
for the case $t_y=1.3 t_x$, see the text 
}
\label{1}
\end{figure}

Plus (minus) sign applies when $m$ and $m'$ are equal (different), and
\begin{equation}
\nonumber
F^{\pm}_{{\bf k},{\bf q}}=
   1 \pm
\frac{
   \left ( \epsilon_{{\bf k}+{\bf q}+\pi_x}-\epsilon_{{\bf k}+{\bf q}}
   \right )
   \left ( \epsilon_{{\bf k}+\pi_x}-\epsilon_{\bf k}
   \right )+ 
   V^2}
{\sqrt{
    \left ( \epsilon_{{\bf k}+{\bf q}+\pi_x}-\epsilon_{{\bf k}+{\bf q}}
    \right )^2 + 
    V^2}
\sqrt{
    \left ( \epsilon_{{\bf k}+\pi_x}-\epsilon_{\bf k}
    \right )^2 + 
    V^2 }}
\end{equation}
\begin{equation}
\Lambda^{m m'}_{{\bf k},{\bf q}}=\frac{1}{4}
   \left( 1-\frac{\epsilon^m_{\bf k} \epsilon^{m'}_{{\bf k}+{\bf q}}+
\Delta^m_{\bf k}\Delta^{m'}_{{\bf k}+{\bf q}}}{E^m_{\bf k} E^{m'}_{{\bf k}+{\bf q}}}
   \right )
\end{equation}
Neutron scattering experiments are affected by the resolution factor of the 
measurement apparatus. In the experiments by Stock {\it et al}\cite{Stock04} 
the resolution factor in ${\bf q}$ space is anisotropic and comparable to the observed incommensurability. In order to compare with their results we convolute  $Im \chi (q,\omega)$ with an
anisotropic gaussian\cite{resolution} 
with standard deviation along directions perpendicular and parallel to the chains $\sigma_x=0.078$ and $\sigma_y=0.058$. We set
$t'=0.15$, $\Delta_{SP}=0.31$, $V=0.6$, and 
$\mu=-0.70$, in units of $t$, corresponding to hole doping $x=0.09$. 
In the following, we assume $t=180$ meV,
then $\Delta_{spin}=0.183$ and $A=0.52$.

\begin{figure}
 \begin{center}
\leavevmode
\epsfxsize=85mm 
\epsfbox{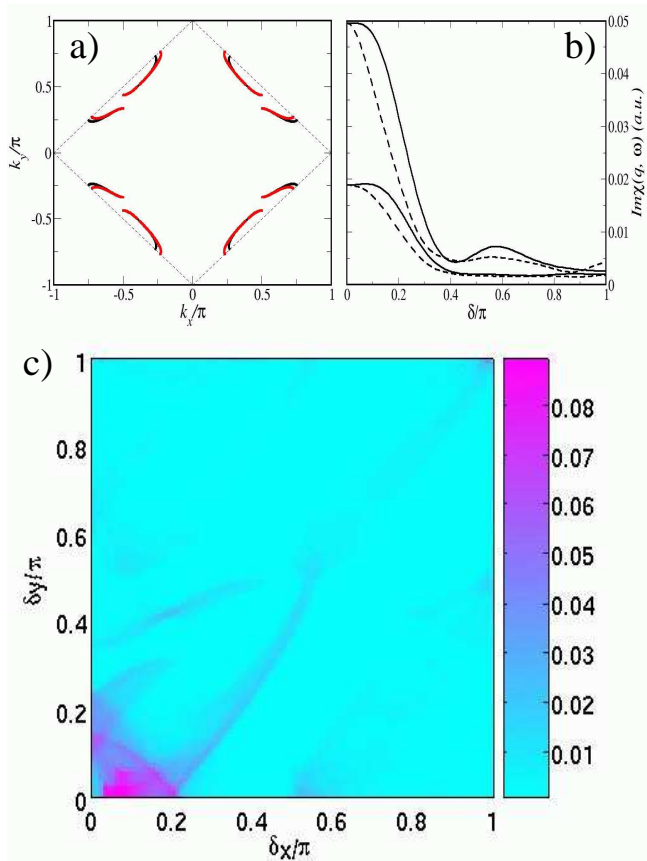}
 \end{center}
\vspace{0mm}
\caption{a) Fermi surface corresponding to the case which includes both 
saddle-point gap anisotropy and ortho-II ordering (red), and only 
ortho-II ordering (black). 
The later one is also plotted in Fig. 1. b) Spin susceptibility at $\omega=0.05$-convoluted with the resolution factor- 
as a function of momentum, measured from $(\pi,\pi)$, along the $(\delta,0)$ 
and $(0,\delta)$, in solid and dashed lines, respectively,  for $\alpha=-0.33$ 
in the normal (top curves) and superconducting state with $\Delta_0=0.1$ (bottom curves). c) Colormap of the spin response for the same parameters as in b) in the normal state, but without including resolution factor effects.}
\label{2}
\end{figure}

We first consider the case $\Delta_0=0$
$\alpha=0$ and analyze the possibility that the observed anisotropy is a consequence of 
the modification of the Fermi surface due to ortho-II ordering. 
The truncated Fermi 
surface obtained with these parameters is shown, in the extended Brillouin 
zone, in Fig. 1a). The breaking of the tetragonal symmetry is clear. 
The magnetic response (not shown)
has two peaks close to $(\pi,\pi)$, along
the directions $(\pi+\delta,\pi)$ and $(\pi,\pi+\delta)$.
The position of these peaks depends on the band parameters and their distance
from $(\pi,\pi)$ increases with doping. 
The spin susceptibility, convoluted with the resolution factor,
along $(\pi+\delta,\pi)$ and $(\pi,\pi+\delta)$ is shown in 
Fig. 1b).  
As the value of $J$ is uncertain, the susceptibility is given
in arbitrary units, but these units are the same throughout. 
In spite of the clear anisotropy
of the Fermi surface, once the resolution factor is included,
the spin response in the directions perpendicular and
parallel to the chains is very similar. We have repeated the calculations for
different Fermi surface parameters. The anisotropy we obtained is very weak or it gives
larger incommensurability in  $(\pi,\pi+\delta)$, contrary to the experiment. 
We have checked the effect which, combined with the ortho-II ordering, could
have the addition of a third nearest neighbor hopping term, the reduction
of the spectral weight along the Fermi surface as the saddle points are approached or the inclusion of finite lifetime of the quasiparticles\cite{lifetime}. 
None of these
effects gives an incommensurability comparable to the observed value.
We conclude that the modification of the Fermi surface due to 
the ortho-II symmetry breaking alone cannot explain the
experimental results.

\begin{figure}
 \begin{center}
\leavevmode
\epsfxsize=85mm 
\epsfbox{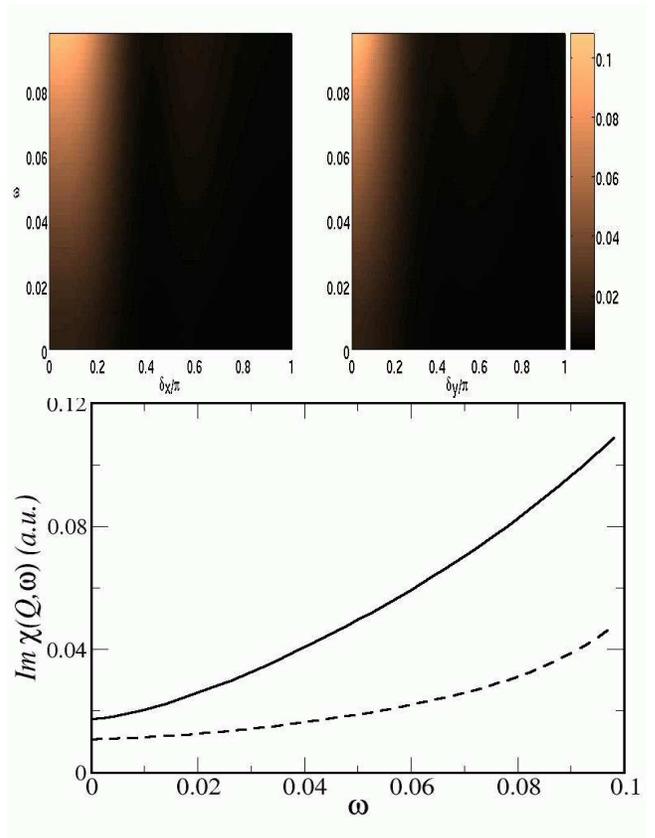}
 \end{center}
\vspace{0mm}
\caption{Dependence of the spin response -convoluted with the resolution factor- on energy for $\alpha=-0.33$. Top figure: color map as a function of
energy and momentum, measured from $(\pi,\pi)$, along $(\delta,0)$ 
(right figure) and $(0,\delta)$ (left figure). Bottom figure: at $(\pi,\pi)$.
Solid and dashed lines correspond to the normal and superconducting states, respectively.}
\label{3}
\end{figure}

The anisotropy in the hopping matrix element has been suggested\cite{Eremin05}, alone or with other effects,
 as the origin of the one-dimensionality of the spin spectrum reported by
Mook {\it et al}\cite{Mook00} in YBa$_2$Cu$_3$O$_{6.6}$.  
The case in which the hopping along $(t_y)$ and perpendicular $(t_x)$ to the chains 
are inequivalent is shown in the inset of Fig. 1 b). We have 
taken $t_y=1.3 t_x=1.3 t$, keeping 
the rest of the parameters as above. With this modification of 
the Fermi surface the tail of the peak at $(\pi,\pi)$ decays more slowly along $(\pi+\delta,\pi)$, than along $(\pi, \pi+\delta)$. But this broadening is different to the one found in ortho-II YBa$_2$Cu$_3$O$_{6.5}$. We have also 
checked the 
effect of a modulated doping, which could follow the ortho-II oxygen ordering,
with no success.

Next we examine the effect of finite $\alpha$. The gap at the AFBZ loses 
its d-wave symmetry and for $\alpha=-0.33$,  
at $(\pi,0)$ it is twice than at $(0,\pi)$. In spite of this large asymmetry, 
the effect on 
the shape of the Fermi arc is not unreasonable, see Fig. 2 a). 
The crossing between the arcs and the AFBZ moves towards $(0,\pi)$ in the first and second quadrants and towards $(0,-\pi)$ in the third and fourth quadrants.
The color map in Fig. 2 c) shows the spin susceptibility at 
$\omega=0.05$, without the 
resolution factor. The inequivalence of the peaks is clear. This anisotropy
remains when the resolution factor is included.
The shape of the peaks along $(\pi+\delta,\pi)$ and $(\pi,\pi+\delta)$ are 
plotted 
in Fig. 2 b),
and compare well with those reported\cite{Stock04}. 
As shown in Fig. 3 (top figure) the 
anisotropy is maintained with increasing energy. As the resonance is 
approached the incommensurability vanishes due to the enhancement factor.
The anisotropy is robust and survives also in the 
superconducting state, see Fig. 2b). 
In this plot it is clear that at low energies, the 
magnetic intensity has been reduced in the superconducting state. 
The suppression in the superconducting state is 
due to the opening of a gap along the arcs.
It is
also evident in Fig. 3 (bottom figure) which plots the susceptibility 
at $(\pi,\pi)$ as
a function of energy. The signal increases with energy, as 
reported experimentally.

In conclusion, we have presented a simple phenomenological model which is able
to account for the main features of the spin scattering in ortho-II 
YBa$_2$Cu$_3$O$_{6.5}$, at low energies, without involving one-dimensionality. The model should be applicable to 
other underdoped cuprates. 
The spin response is larger close to $(\pi,\pi)$, but sensitive to the 
parameters which determine the Fermi surface. It is characterized by two 
peaks along the Cu-O axis. Below $T_c$ 
the opening of the superconducting gap on the arcs produces a
reorganization of 
the spectral 
weight at low energies. The 
superconducting state can influence the spin susceptibility also
through a better definition of the quasiparticles. The model is intended for 
low energies in the underdoped regime. The high energy part 
observed experimentally, including the resonance, is modelled with a
collective triplet mode .
Recently, Tranquada {\it et al}\cite{Tranquada04} 
has suggested that the high energy mode is
due to disordered bond-stripes, which behave as weakly coupled two-leg 
ladders. 
However in our view the ring form of the dispersion at high energies suggests a 2-dimensional character for the spin liquid.
Our model can be seen as a phenomelogical extension for the pseudogap phase of the RVB (resonant valence bond) theory, whose success was recently reviewed\cite{RVB}.

The anisotropy of the spin response found in Ortho-II YBa$_2$Cu$_3$O$_{6.5}$ is ascribed to an asymmetric deformation
of the Fermi arcs close to the boundary of the AFBZ. We found that such a deformation could originate from an anisotropic $\Delta_{SL}$ -the energy gap that truncates the Fermi surface in the normal pseudogap phase.
A substantial anisotropy of $\Delta_{SL}$  was required to explain the 
experiment. 
The origin of this anisotropy remains to be an open question, but could be 
related to the chain ordering.

We thank W.J.L. Buyers, A. Chubukov, B. Fauqu\'e, A. L\"auchli, S. Pailh\'es, 
H. Ronnow, C. Ruegg, A. Schnyder and B. Valenzuela for useful discussions. 
Financial support from NCCR MaNEP of the Swiss Nationalfonds and the 
Spanish Science and Education Ministry through Ram\'on y Cajal contract 
is acknowledged.


\begin{thebibliography}{99}
\bibitem{resnote} This enhancement is intermediate between the one resulting
from Fermi-Golden rule and the one obtained from an RPA-coupling between the
bare particle-hole susceptibility and the high-energy mode.
\bibitem{resolution}We have neglected finite resolution in energy. Due to the robustness of the anisotropy with energy we do not
expect any effect of the finite resolution in energy on  the anisotropy found.
\end{thebibliography}
\end{document}